\documentclass[conference]{IEEEtran}
\ifCLASSINFOpdf
\usepackage[pdftex]{graphicx}
\else
\fi
\usepackage[cmex10]{amsmath}
\usepackage{amsfonts}
\usepackage{mathtools}
\usepackage{url}
\usepackage{citesort}
\usepackage[utf8]{inputenc}  
\usepackage[T1]{fontenc}
\newcommand{\defeq}{\vcentcolon=}

\begin{document}

\title{Replica analysis and approximate message passing decoder
for superposition codes}

\author{\IEEEauthorblockN{Jean Barbier and Florent Krzakala}
\IEEEauthorblockA{Laboratoire de Physique Statistique, CNRS UMR 8550, Ecole Normale
Supérieure, 24 rue Lhomond, 75005 Paris, France\\
Université P. et M. Curie, UPMC Paris and CNRS UMR 7083, 10 rue Vauquelin, 75005 Paris, France \\
Email: jean.barbier@ens.fr, florent.krzakala@ens.fr}
}

\maketitle

\begin{abstract}
 Superposition codes are efficient for the additive white gaussian
  noise channel. We provide here a replica analysis of the
  performances of these codes for large signals. We also consider a
  Bayesian approximate message passing decoder based on a
  belief-propagation approach, and discuss its performance using the
  density evolution technic. Our main findings are 1) for the sizes we
  can access, the message-passing decoder outperforms other decoders
  studied in the literature 2) its performance is limited by a sharp
  phase transition and 3) while these codes reach capacity as $B$ (a
  crucial parameter in the code) increases, the performance of the
  message passing decoder worsen as the phase transition goes to lower
  rates.
\end{abstract}

\IEEEpeerreviewmaketitle
\vspace{0.5cm}
Superposition coding is a scheme for error-correction over the
Additive White Gaussian Noise (AWGN) channel where a codeword $\tilde
Y$ is a sparse linear superposition of a random independent and identically distributed (i.i.d) matrix $F$. Even
though it has been shown in \cite{barron2010sparse,joseph2012least}
that these codes (in a proper limit) are reliable ---up to
exponentially small error--- up to capacity, the performance of the
overall scheme is highly dependent on the decoder efficiency. In
particular, \cite{barron2010sparse} have proposed an iterative
algorithm called {\it adaptive successive decoder}, later on improved in \cite{barron2012high}. In the present
work, we expose another kind of iterative procedure, based on a
Bayesian approach combined with a Belief-Propagation (BP) type
algorithm, using technics that have been originally developed for
compressed-sensing: the so-called Approximate Message Passing
algorithm (AMP)
\cite{DonohoMaleki09,BayatiMontanari10,Rangan10b,KrzakalaPRX2012}. Much
in the same way BP is used in the context of Low Density Parity Check
(LDPC) codes \cite{RichardsonUrbanke08}, the AMP approach combines the
knowledge of the noise and signal statistics with the powerful inference
capabilities of BP. A second contribution we provide is the
computation of the performance of these codes under optimal decoding
using the (non-rigorous) replica method
\cite{MezardParisi87b,MezardMontanari09}. The approach is deeply
related to what has been previously applied to LDPC codes.

This contribution is organized as follow: In Sec.~\ref{sec:Super}, we
briefly present superposition codes and introduce the notations.  The
replica analysis is performed in Sec.~\ref{sec:replica}.  In
Sec.~\ref{sec:AMP} we describe the AMP algorithm and study its
performance by the Density Evolution \cite{BayatiMontanari10} (DE)
technic in Sec.~\ref{sec:DE}. Finally, Sec.~\ref{sec:numerical}
presents a numerical study of the performances for finite size
signals.

\section{Superposition codes}
\label{sec:Super}
We refer the reader to the original papers \cite{barron2010sparse,barron2011analysis,joseph2012least} for a full description of superposition codes. The
message to be transmitted is a string $S=[s_1,s_2,..,s_L]$ where each
$s_i \in \{1,2,..,B\}$. It is converted onto a binary string $X$ of
dimension $N=LB$ where in each of the $L$ sections of size $B$, there
is a {\it unique} value $\neq$ 0 at the position corresponding to the
state of the associated variable in $S$ (using a power of $2$ for $B$ ensures
that this step is trivial). One then introduce the coding matrix $F$ of dimensions $M \times N$ ($M < N$) whose elements are i.i.d Gaussian distributed with mean $0$ and
variance $\sigma_F^2=1/L$. The codeword (there are $B^L$ of them) reads: $\tilde Y = FX$.

We choose the scaling of $F$ such that $\tilde Y$ has a unit constant power
$P \defeq 1/M \sum_{\mu = 1}^M \tilde y_\mu^2 = 1$, the only relevant parameter is thus the signal to noise ratio ${\rm{snr}}\defeq 1/{\sigma^2}$. The dimension of
$\tilde Y$ is $M \defeq N\log_2(B)/(RB)$ where $R$ is the
coding rate in bits per channel use. The noisy output of
the AWGN channel with capacity
$C=\log_2 \left(1+{\rm{snr}}\right)/2$ reads
\begin{equation}
Y \defeq \tilde Y + \xi = FX + \xi, \ \xi_\mu \sim \mathcal{N}_{0, \sigma^2} \ \forall \mu \in \{1,...,M\}.
\end{equation}
where $\mathcal{N}_{m, \sigma^2}$ is a Gaussian distribution with mean $m$ and variance $\sigma^2$. Now comes the crucial point that underline our approach: Consider that
each section $l \in \{1,2,...,L\}$ of $B$ variables in $X$ is {\it one single}
$B$-dimensional ($B$-d) variable on which we have the prior information that
it is zero in all dimensions {\it but} one. In other words, instead of
dealing with a $N$-d vector $X$ with elements $x_i$, we deal
with a $L$-d vector $\overline X$ whose elements $\overline
x_l$ are $B$-d vectors ($\overline x_l$ thus contains the
information on the $l^{th}$ section of variables in $X$). In this
setting, decoding $\overline X$ from the knowledge of $Y$ and the
dictionary $F$ is a (multidimensional) linear estimation problem with
element-wise prior information on the signal. This is exactly the kind
of problem considered in the Bayesian approach to compressed sensing
\cite{Rangan10b,montanari2012graphical,KrzakalaPRX2012,KrzakalaMezard12}.
\section{Replica analysis}
\label{sec:replica}
\begin{figure}[!t]\hspace{0cm}
\includegraphics[width=1\linewidth]{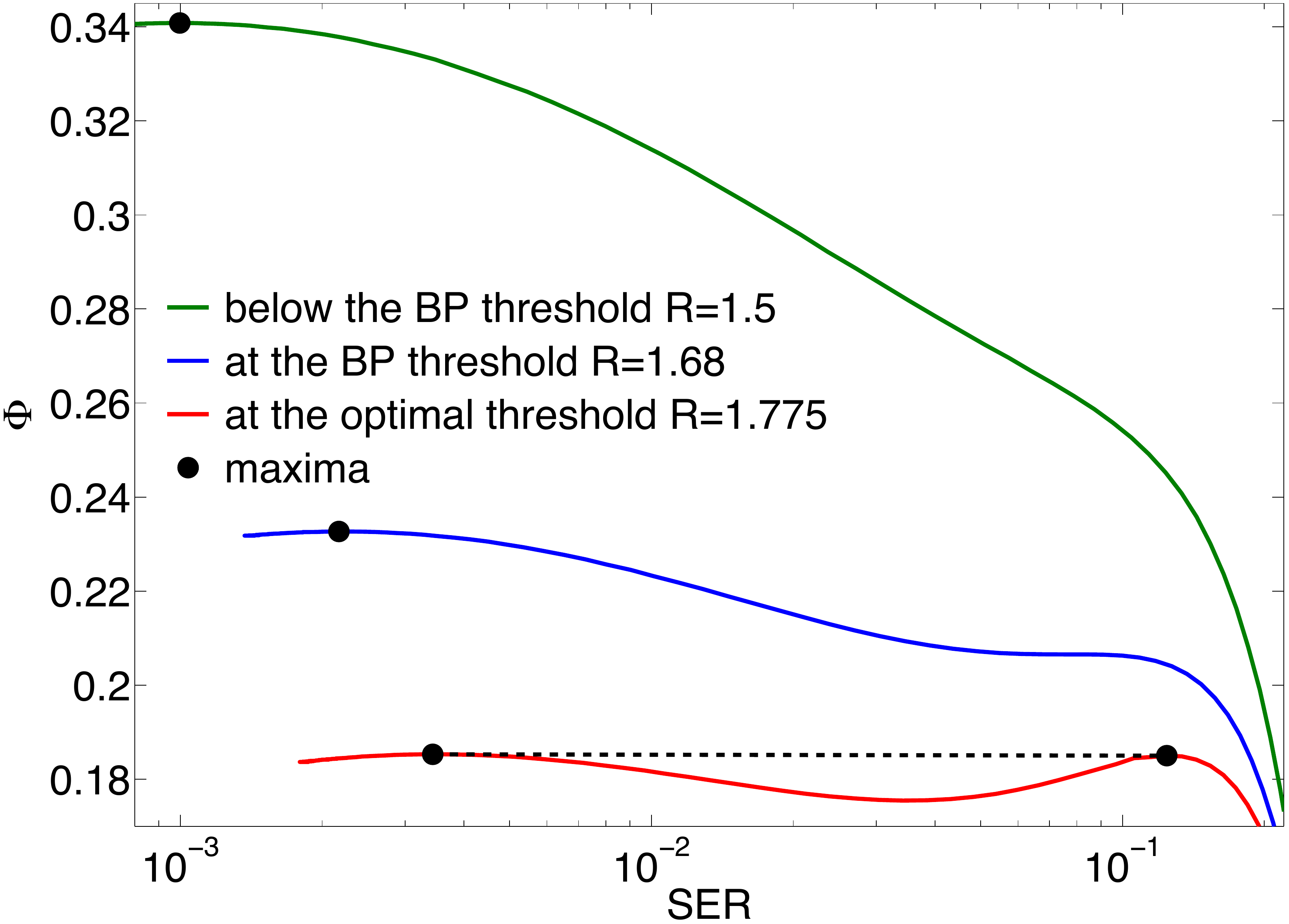}
\caption{The free-entropy (or potential) $\Phi$ as a function of the
section-error rate ${\rm{SER}}$ for $B = 2$ and ${\rm{snr}}=30$. The
global maximum of the curves allows to identify the typical optimal ${\rm{SER^*}}$ of the
codes. The curves are obtained by numerical integration of
(\ref{eqFreeEnt}). Here for rates larger than $R>1.775$, the
optimal ${\rm{SER^*}}$ jumps from a low value to a large $O(1)$ one (red
curve). This defines the maximum possible rate (to compare here to
$C=2.4771$) below which acceptable performance can be obtained. For
$R<1.775$, the ${\rm{SER^*}}$ is much lower (and decay with $R$). The AMP
decoder described in Sec.~\ref{sec:AMP} allows to perform an ascent
of this function. As long as the maximum is unique (i.e. for
$R<1.68$, see blue curve), it will be able to achieve the predicted optimal 
performance in the large size limit.}
\label{freeEnt}
\end{figure}
Let us first access the performances of these codes when $B$ is finite
in the large $L\!\to\!\infty$ limit. We proceed as in the LDPC case
\cite{RichardsonUrbanke08} and define the "potential" function (or
free-entropy \cite{MezardMontanari09}) at fixed $B$ as
$\Phi_B\!\defeq\!\langle \log Z(Y,F) \rangle_{X,F,\xi}$ where $\langle
\rangle_u$ denotes the average over the random variable $u$ and
$Z(Y,F)$ is the normalization of the probability measure of the estimator $\hat X$:
\begin{equation}
P_{Y,F}(\hat X) = \frac{1}{Z(Y,F)} \prod_{l=1}^L P_0(\hat{\overline x}_l) \prod_{\mu=1}^M e^{-\frac{{\rm{snr}}}{2}(y_\mu-\sum_{i=1}^N F_{\mu i} \hat x_i)^2}
\label{probMeasure}
\end{equation}
where $\hat {\overline x}_l =[\hat x_1, \hat x_2,...,\hat x_B]_l$ is the
$l^{th}$ $B$-d variable of the estimator (i.e one of its
$L$ sections as for $\overline X$).
Our computation is based on the replica method, an heuristic coming
from the statistical physics of disordered systems that relies on the
identity $\langle \log Z(u) \rangle_u = \lim_{n\!\to\!0} (\langle Z(u)^n
\rangle_u-1)/n$ to compute averages of logarithms, and that has been
applied very often to coding theory \cite{MezardMontanari09}, sparse
estimation
\cite{Tanaka02,KabashimaWadayama09,KrzakalaPRX2012,KrzakalaMezard12}
or optimization problems \cite{MezardParisi87b} where it has been
later shown to rigorously give the correct answer. We shall not give
a detailed description of our computation that in fact follows the one
given in \cite{KrzakalaMezard12} for compressed sensing almost step by
step. The only difference is again that we are dealing here with
$B$-d variables $\overline x_l$ whose components takes
binary value $\in \{0, {\rm{val}}_l\}$, and only one of the $B$ $x_i$'s in $\overline
x_l$ is non zero. Any value ${\rm{val}}_l \neq 0$ is a possible choice
(such as the exponentially distributed ${\rm{val}}_l$ of
\cite{barron2010sparse,barron2011analysis}) but we shall here restrict ourselves to
the simplest case where ${\rm{val}}_l = 1 \ \forall \ l \in \{1,...,L\}$. We have observed
empirically that this seems to be an efficient distribution for the
algorithm we describe in Sec.~\ref{sec:AMP}.

The estimator we are interested in is the section-error rate
(${\rm{SER}}$), the fraction of incorrectly decoded sections:
\begin{align}
\label{def:SER1}
&{\rm{SER}} \defeq \frac{1}{L} \sum_{l=1}^L \mathbb{I} \left({\rm{argmax}} [P(\hat{\overline x}_l)] \ne \overline x_l \right)\\
\label{def:argmax}
&{\rm{argmax}} [P(\hat{\overline x}_l)] = [\hat x_1, \hat x_2,...,\hat x_B]_l \\ 
&\hat x_i \defeq \mathbb{I} \left(a_i > a_j \ \forall j \in l: j\neq i \right)
\label{def:SER2}
\end{align}
where $a_i$ is the posterior average with respect to
(\ref{probMeasure}) of the $1$-d variable $x_i$,
$P(\hat{\overline x}_l)$ is the marginal probability of section $l$ (i.e
the joint probability $P(\{\hat x_i : i \in l\})$) and $\mathbb{I}()$
is the indicator function. Most of our results, however, can be
expressed more conveniently as function of a different estimator that
we shall denote as the biased MSE $\tilde E$:
\begin{equation}
\label{def:Etilde}
\tilde E \defeq \frac{1}{N}\sum_{i=1}^N (a_i - x_i)^2
\end{equation}
Both quantities behave similarly in our computations. In fact, the
replica and DE analysis show that the ${\rm{SER}}$ can be
computed directly from the value of $\tilde E$ (see Sec.~\ref{sec:DE},
and in particular eq.(\ref{DE_E})).

The main interest of the potential is that the actual typical value
$\tilde E^*$ (and thus the typical value ${\rm{SER^*}}$) for a given ensemble
of codes in the large signal size limit is obtained by maximizing it
so that $\tilde E^*={\rm{argmax}} [\Phi(\tilde E)]$. After
computation and defining a rescaled biased MSE $E \defeq \tilde E B$ one can show that for a given $B$ and up to constants,
the potential reads: 
\begin{eqnarray}
&\Phi_B(E) = -\frac{\log_2(B)}{2R} \left(\log(1/{\rm{snr}} + E) + \frac{1 - E}{1/{\rm{snr}} + E} \right) \nonumber\\
&+ \int \mathcal{D}\overline z \log\left( e^{\frac{1}{2\Sigma(E)^2} + \frac{z_1}{\Sigma(E)} } + \sum_{i = 2}^B e^{-\frac{1}{2\Sigma(E)^2} + \frac{z_i}{\Sigma(E)} }\right)
\label{eqFreeEnt}
\end{eqnarray}
where $\Sigma(E)\defeq \sqrt{(1/{\rm{snr}} + E)R/\log_2(B)}$ and
$\mathcal{D}\overline z \defeq \prod_{i=1}^B \mathcal{D} z_i$ (with $\mathcal{D}
z_i$ being a centered Gaussian measure of unit variance). Using the mapping from
$\tilde E$ to ${\rm{SER}}(\tilde E)$ given by eq.(\ref{DE_E}) we can compute numerically
$\Phi_B({\rm{SER}})$. Exemples are shown in Fig.~\ref{freeEnt} in the $B\!=\!2$
case for different rates. 

Interestingly, the potential behaves very similarly to the one of LDPC
codes so that we observed the very same phenomenology. For large
enough rates, it develops a "two-maxima shapes" with a high and low
error maxima. The low error maxima dominates as long as it is the
higher one, i.e statistically dominant. When the two maxima of
$\Phi_B$ have same height (red curve on Fig.~\ref{freeEnt}) the
reconstruction capacities become extremely poor: this mark what we
shall refer to as the "optimal threshold", as it represents the point
until which the code has good performance under optimal decoding.

\begin{figure}[!t]\hspace{0cm}
\includegraphics[width=1\linewidth]{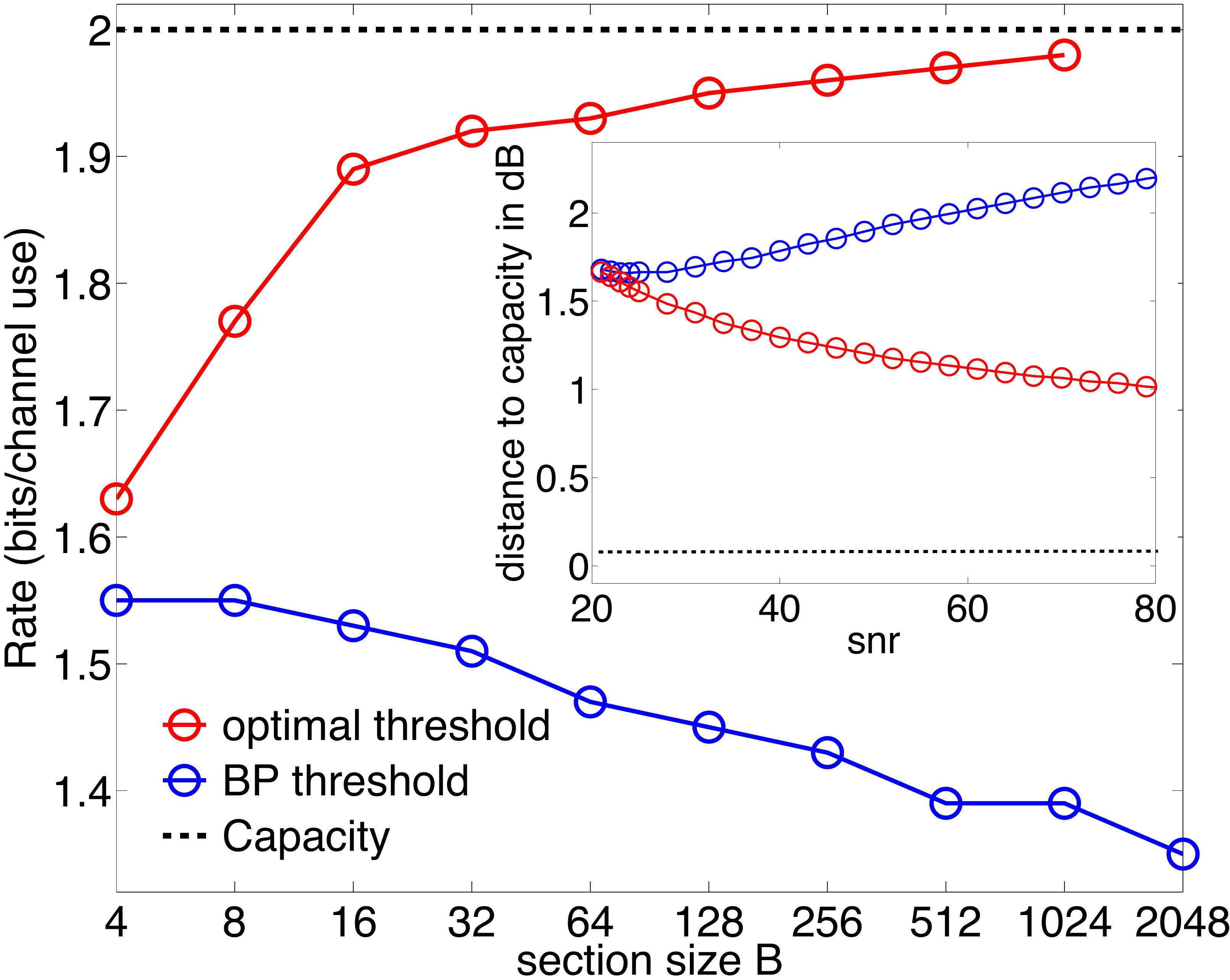}
\caption{Main Fig: Phase diagram for ${\rm{snr}}=15$ changing $B$. The two transition
lines (blue and red) are respectively the BP threshold and the
optimal one (see Fig.~\ref{freeEnt}). At the optimal transition the
${\rm{SER^*}}$ drops from $O(1)$ values to much lower
ones. Below the BP threshold, the optimal ${\rm{SER^*}}$ is
achievable using the AMP decoder described in
Sec.~\ref{sec:AMP}. Notice that while the optimal threshold tends to
the capacity $C$ as $B$ increases, the BP threshold unfortunately
decays when $B$ increases. Inset: Phase diagram with the same transitions
plotted in distance in dB from capacity at fixed $B\!=\!2$ and changing
the ${\rm{snr}}$. Below ${\rm{snr}}\approx 20$, there are no more
transition and the ${\rm{SER}}$ decreases continuously with the
rate. Curves are obtained by Monte Carlo integration with sample size of
$10^6$ of eq.(\ref{eqFreeEnt}) and eq.(\ref{DE_E}) for $B>2$ and
direct numerical integration for $B\!=\!2$.} 
\label{phaseDiag}
\end{figure}

It is well known in LDPC codes that the rate for which the second
maxima appears (blue curve on Fig.~\ref{freeEnt}) corresponds to the
moment where the BP decoder does not correspond anymore to the optimal
one. We shall see that this remains the case here and that only in the
one-maxima region (green curve on Fig.~\ref{freeEnt}), our AMP decoder
will match asymptotically the optimal
performance. Fig.~\ref{phaseDiag} represents these two thresholds in
the limit $L\!\to\!\infty$ as a function of the section size $B$ for
fixed ${\rm{snr}}=15$. We shall come back to the BP performances, but
let us first notice here that (i) as $B$ increases, the optimal
threshold approaches the Shannon capacity (and in fact the
corresponding value of ${\rm{SER}}$ goes to zero, see
Fig~\ref{optSER}). However, (ii) the rate at which we expect message
passing algorithms to converge to the optimal value unfortunately {\it
  decays} as $B$ gets larger.  These two observations are in agreament
with the $B\!\to\!\infty$ limit, performed by replica analysis: The
optimal threshold matches the capacity, as predicted in
\cite{barron2010sparse} and the rate at the BP threshold is
asymptotically given by:
\begin{equation}
R_{BP} \underset{L\to\infty}{=} [(1/\rm{snr} + 1)2\log2]^{-1}
\end{equation}
\nopagebreak

\vspace{-1cm}
\section{Approximate Message Passing algorithm}
\label{sec:AMP}
We shall now consider a BP type iterative algorithm to estimate the
joint probability $P(\hat{\overline x}_l)$ of one section, still in the case where ${\rm{val}}_l=1\ \forall \ l$. 
Following the derivation of the AMP in
\cite{KrzakalaMezard12}, it follows from the parametrization:
\begin{align}
P(\hat{\overline x}_l) &= \frac{P_0(\hat{\overline x}_l)}{z} \prod_{\{i \in l\}}^B m(\hat x_i)\label{jointProba}\\
z &= \sum_{\{i \in l\}}^B e^{-\frac{(1-R_i)^2}{2\Sigma_i^2}}\prod_{j\neq i}^{B-1} e^{-\frac{R_j^2}{2\Sigma_j^2}} \\
P_0(\hat{\overline x}_l) &= \frac{1}{B} \sum_{\{i \in l\}}^B \delta(\hat x_i-1)\prod_{j\neq i}^{B-1}\delta(\hat x_j)
\end{align}
where $P_0()$ is the prior that matches the signal distribution, eq.(\ref{probMeasure}). This matching condition is called Nishimori
condition in statistical physics
\cite{NishimoriBook,KrzakalaMezard12}. The set $\{m(\hat x_i) \sim
\mathcal{N}_{R_i, \Sigma^2_i} \}$ are Gaussian distributed, with
moments iteratively computed by the AMP. Marginalizing over $\{\hat
x_j \in \hat{\overline x}_l \neq \hat x_i\}$ we get after
simplification the posterior average $a_i^t$ and variance $v_i^t$ of
$\hat x_i$ at step t of the algorithm:
\begin{align}
a_i^t &\defeq f_{a_i}(\{\Sigma_j^t,R_j^t\}_{j\in l}) = \frac{e^{-\frac{1-2R_i^t}{2(\Sigma_i^t)^2}}}{\sum_{\{j \in l\}}^B e^{{-\frac{1-2R_j^t}{2(\Sigma_j^t)^2}}}} \\
v_i^t &\defeq f_{c_i}(\{\Sigma_j^t,R_j^t\}_{j\in l}) =  a_i^t (1- a_i^t ) 
\end{align}

AMP allows to obtain the estimate at time $t$ given the
parametrization (\ref{jointProba}) as follows (see again \cite{KrzakalaMezard12}):
\begin{equation*}
\boxed{
\begin{aligned}
V^{t+1}_\mu &= \sum_i F_{\mu i}^{2} v_i^t \\
\omega^{t+1}_\mu  &= \sum_i F_{\mu i} a^t_i - (y_\mu-\omega^t_\mu)\frac{V^{t+1}_\mu}{1/{\rm{snr}}+V^{t}_\mu} \ \ \\
(\Sigma^{t+1}_i)^2 &= \left[ \sum_\mu \frac{F^2_{\mu i}}{1/{\rm{snr}}+V^{t+1}_\mu} \right]^{-1} \\
R^{t+1}_i &= a^t_i + (\Sigma^{t+1}_i)^2 \sum_\mu F_{\mu i} \frac{(y_\mu - \omega^{t+1}_\mu)}{1/{\rm{snr}}+V^{t+1}_\mu} \ \ \\
a^{t+1}_i &= f_{a_i}\left(\{\Sigma_j^{t+1},R_j^{t+1}\}_{j\in l}\right) \\
v^{t+1}_i &= f_{c_i}\left(\{\Sigma_j^{t+1},R_j^{t+1}\}_{j\in l}\right)
\end{aligned}}
\end{equation*}
Here only the functions $f_{a_i}()$ and $f_{c_i}()$ depend in an explicit
way on the signal model $P_0()$. A suitable initialization for the
quantities is ($a_i^{t=0}\!=\!0$, $v_i^{t=0}\!=\!1/(B{\rm{snr}})
$, $\omega_\mu^{t=0}\!=\!y_\mu$). Once the convergence of the iterative
equations is reached, one get the posterior average of the
$i^{th}$ signal component given by $a^{t}_i$ and the
${\rm{SER}}^t$ as well using ${\rm{argmax}}[P(\hat{\overline x}_l^t)]$, eq.(\ref{def:argmax}).

\section{Asymptotic analysis of AMP}
\label{sec:DE}
\begin{figure}[!t]\hspace{0cm}
\includegraphics[width=1\linewidth]{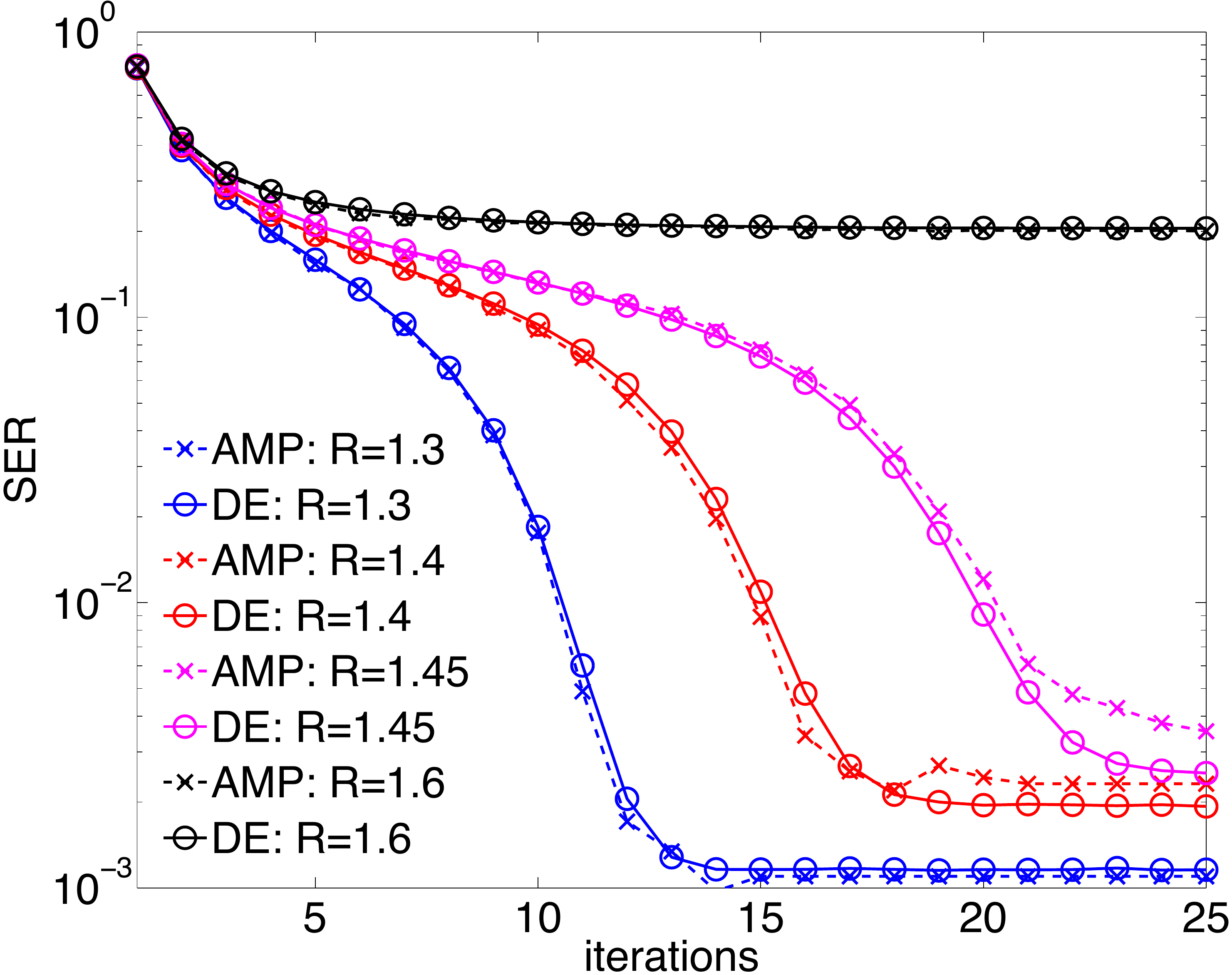}
\caption{The density evolution predictions of the ${\rm{SER}}^t$ for $t\ge1$
(${\rm{SER}}^{t=0}=1$) compared to the actual one of the algorithm
for ${\rm{snr}}=15$ and different rates (to be compared to the BP
threshold at $R_{BP}=1.55$) and a section size $B\!=\!4$. DE is
computed by a Monte Carlo technique with a sample size of $10^7$ and the signal
size for AMP is $N=2^{15}$. The figure shows how close is the prediction from the true behavior for finite sizes.}
\label{illustrDE}
\end{figure}
The dynamics of the ${\rm{SER}}^t$ under the AMP recursions can be
described in the large signal size ($N\!\to\!\infty$,
$L\!\to\!\infty$, $B\!=\!N/L$ fixed) using the DE
technic \cite{BayatiMontanari10}. It consists in the statistical
analysis of the quantities $(a_i^t, v_i^t)$, which is based on the
fact that the product in (\ref{jointProba}) becomes Gaussian
distributed (by central-limit theorem) as $L\to\infty$.

As for the replica analysis, our derivation follows
\cite{KrzakalaMezard12} with the only difference that we are dealing
with $B$-d variables. Defining the effective variance
$\Sigma^t$ and the rescaled biased MSE as in (\ref{def:Etilde}) which now
depends on the iteration step:
\begin{equation}
\Sigma^t \defeq \sqrt{(1/{\rm{snr}} + E^t)R/\log_2(B)}, \ E^t \defeq \frac{1}{L}\sum_{i=1}^N (a_i^t - x_i)^2
\end{equation}
then the rescaled biased MSE follows the following iteration in the $L\!\to\!\infty$ limit:
\begin{equation}
E^t\!\!=\!\!\int\!\!\mathcal{D} \overline z \left([f_{a_1}^{(1)}(\Sigma^{t-1},\overline z)-1]^{2}+(B-1)[f_{a_{1,2}}^{(0)}(\Sigma^{t-1},\overline z)]^{2}\right)
\end{equation}
where:
\begin{align}
f_{a_i}^{(1)}(\Sigma,\overline z)  &\defeq \left[1 + e^{-\frac{1}{\Sigma^2}}  \!\!\!\!\!\!\!\! \sum_{\{1\le j\le B : j\ne i\}} \!\!\!\!\!\!\!\!e^{\frac{z_j-z_i}{\Sigma}} \right]^{-1}  \\
f_{a_{i,j}}^{(0)}(\Sigma,\overline z)  &\defeq \left[1 + e^{\frac{1}{\Sigma^2} + \frac{z_j - z_i}{\Sigma}}+ \!\!\!\!\!\!\!\!\sum_{\{1\le k\le B : k \ne i,j\} } \!\!\!\!\!\!\!\!\!\! e^{\frac{z_k-z_i}{\Sigma}} \right]^{-1}
\end{align}
In this approach, there is a one to one correspondance from the value
of the biased MSE to the ${\rm{SER}}$ thanks to the mapping:
\begin{equation}
{\rm{SER}}^{t} \!\!=\! \! \int \! \mathcal{D}\overline z \ \mathbb{I}\left(\exists \ j \in \{2,...,B\} : f_{a_{j,1}}^{(0)}(\Sigma^t,\overline z) > f_{a_1}^{(1)}(\Sigma^t,\overline z) \right)
\label{DE_E}
\end{equation}
The fixed-point of this set of equations for a given $B$, starting
from $E^{t=0}\!=\!1$, allows to compute the asymptotic
${\rm{SER}}$ reached after convergence. Fig.~\ref{illustrDE}
compares the DE prediction of the ${\rm{SER}}^t$ versus the algorithm
dynamics in the $B\!=\!4$ case for different rates. We see how close is
the DE asymptotic theory from the true behavior even for small signals.

Finally, let us now comment on the classical duality between DE
results and the replica potential: DE equations are nothing but
the fixed point iterations of the replica potential
\cite{MezardMontanari09}. As discussed in the Sec.~\ref{sec:replica} we
thus now can identify the BP threshold studied in
Sec.~\ref{sec:replica} as the moment where the DE iterations remain
confined in high ${\rm{SER}}$ values.

\begin{figure}[!t]\hspace{0cm}
\includegraphics[width=1\linewidth]{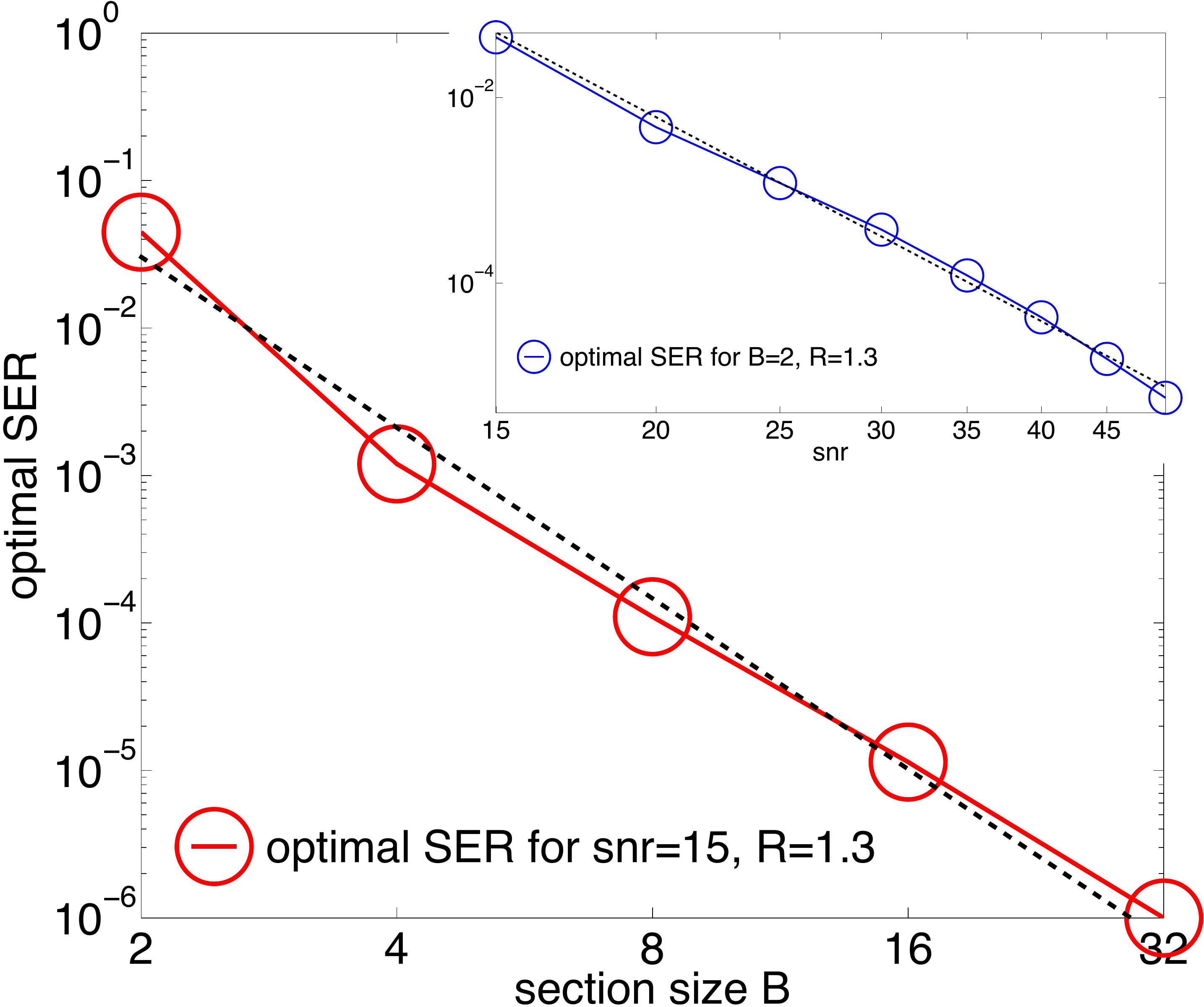}
\caption{Density Evolution prediction of the optimal ${\rm{SER^*}}$
at fixed rate $R=1.3$. The red curve is function of $B$ at fixed
${\rm{snr}}=15$, the blue one at fixed $B\!=\!2$ as a function of the
${\rm{snr}}$. The plots are in double logarithmic scale. The line is a
power law fit to guide the eyes ( ${\rm{SER^*}}(B) \propto
B^{-3.76}$ and ${\rm{SER^*}}({\rm{snr}}) \propto
{\rm{snr}}^{-7.25}$).}
\label{optSER}
\end{figure}

What have we learned from this analysis? The main point is
stressed in Fig.~\ref{phaseDiag}: the BP threshold goes
to larger distance from capacity as $B$ increases. This is unfortunate
since, as shown in Fig.~\ref{optSER}, the optimal ${\rm{SER^*}}$ decays
drastically when $B$ is increased. Nevertheless, AMP alone is not able
to reach optimal performance for high rates.
\section{Numerical experiments}
\begin{figure}[!t]\hspace{0cm}
\includegraphics[width=1\linewidth]{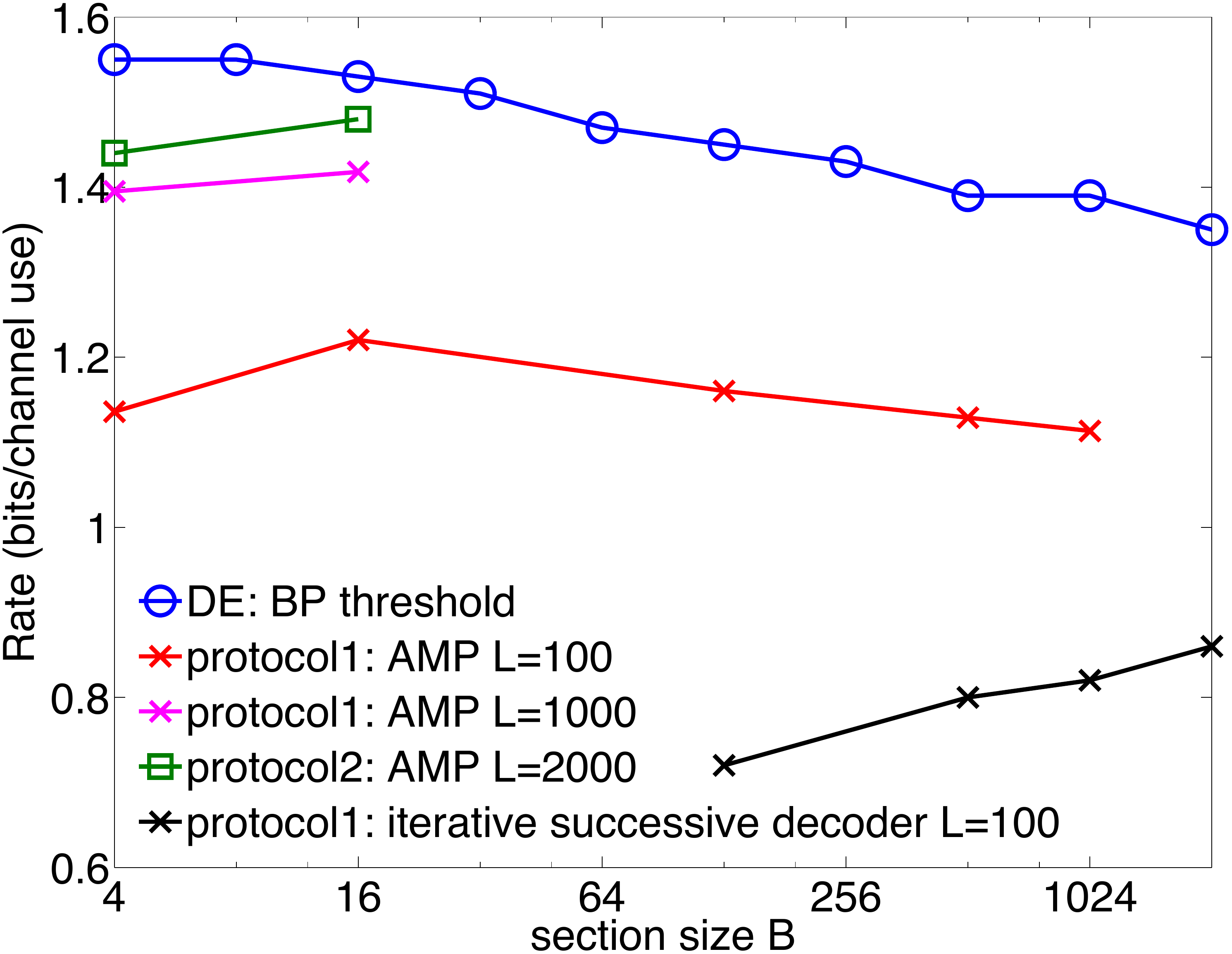}
\caption{Numerical experiment at finite limited sizes $L$ for
${\rm{snr}}=15$. The blue line is the BP threshold rate predicted by
the density evolution technic (to be compared to $C=2$, see
Fig.~\ref{phaseDiag}) and marks the limit of efficiency of AMP for
large signals. Here we limit ourselves to small signals in order to
compare with other results in the literature.  All the curves with
crosses are results of the following protocol (protocol 1): decode
$10^4$ random instances and identify the transition line between a
phase where the probability $p_\epsilon$ to have a
${\rm{SER}}>10^{-1}$ is $p_\epsilon<10^{-3}$ (below the line) from a
phase where $p_\epsilon\ge10^{-3}$ (more than 9 instances have
failed over the $10^4$ ones). The squared line is the result of the
second protocol (protocol 2): at the first try, we require
${\rm{SER}}<10^{-1}$ for $10$ random instances in a row below the
line, not above. Note that in our experiments ${\rm{SER}}<10^{-1}$
essentially means ${\rm{SER}}=0$ at these sizes. We compare our results
with the iterative successive decoder of
\cite{barron2010sparse,barron2011analysis} in protocol 1 where the
number of sections $L=100$ (note that these data, taken from
\cite{barron2010sparse,barron2011analysis}, have been generated
with a code with an exponential signal distribution rather than the
$\{0,1\}$ we used here in order to increase the quality of the
reconstruction). Compared with the red curve (AMP with the same
value of $L$) the better quality of AMP reconstruction is
clear. For both the experimental and asymptotic performances of AMP,
the maximum number of iterations of the algorithm is arbitrarily fixed
to $50$.}
\label{phaseDiagSnr15}
\end{figure}

The previous sections were asymptotic studies of the AMP performances. In this
section we study numerically the influence of finite size effects over
the performance of superposition codes combined with the AMP
decoder. Fig.~\ref{phaseDiagSnr15} summarizes our findings. 
We have limited ourselves to small sizes and performed some numerical
protocols to test the iterative AMP decoding. We checked that not only the Bayesian decoder
does follow the density evolution, but it works
very well for limited size as well, at least far enough from the BP
threshold. In fact, when the rate is below the threshold, the
decoding is usually perfect and is found to reach with high
probability ${\rm{SER}}\!=\!0$. This is due to the fact that in order to
observe an ${\rm{SER}} \in O(\epsilon)$, there must be at least
$L\approx 1/\epsilon$ sections which is not the case for small
signals, when the asymptotic ${\rm{SER}}$ is very small. Even for
reasonably large sizes allowing to get close to the asymptotic
performances (such as $L\!=\!2000$, green curve on
Fig.~\ref{phaseDiagSnr15}), AMP just take seconds to decode on a
Matlab code. 

To confirm that the relevant parameter with respect to the finite size
effects is $L$ and not $N$, we made the following experiment (protocol
1 in Fig.~\ref{phaseDiagSnr15}) fixing $N$ and varying $B$ (thus $L$
decreases as $B$ increases) and observed that the performances worsen
rapidly as $B$ increases.

Finally we compared the efficiency of AMP to the iterative successive
decoder introduced by Baron and Joseph \cite{barron2010sparse}. This
decoder has been shown to be capacity achieving in the large $B$ limit
for a slightly different version of the superposition codes, when an
exponential distribution of the signal entries (see
Sec.~\ref{sec:replica}) is used instead of the $\{0,1\}$ distribution
we have been studying. However, the finite $B$ corrections to the
performance appear to be quite severe. We have compared the
performance of our approach (with the $\{0,1\}$ distribution) with the
result obtained in \cite{barron2010sparse} (see again
Fig.~\ref{phaseDiagSnr15}) and found that at the values of $B$ that
one can consider in a computer, AMP appears to be superior in
performance (though it will eventually be outperformed it in the limit
of very large values of $B$). This remains the case even when compared
to the most recent developements \cite{barron2012high}. It would be
interesting to actually find what is the best possible signal
distribution to further optimize the performances of AMP
reconstruction.
\label{sec:numerical}

\section{Perspectives}
\label{sec:conclusion}

We have provided here a replica analysis of superposition codes and a
Bayesian approximate message passing decoder based on a
belief-propagation like approach that we have analyzed using the
density evolution technic. We have found a behavior very similar to
those found in LDPC codes. In particular, our AMP decoder seems to
have good performances, but is limited to region far from optimal
decoding. We have released a Matlab implementation of our decoder at
\url{https://github.com/jeanbarbier/BPCS_common}.

There are a number of possible continuations of the present work, in
the spirit of recent developments in compressed sensing. Superposition
codes which lies on the fuzzy separation between coding and compressed sensing
theories are in fact a natural venue for these ideas.  The most
natural one is to use the spatial coupling approach
\cite{KudekarRichardson12,KudekarRichardson10,YedlaJian12} in order to
design codes with exact optimal performances but for which AMP will be
able to achieve them for all rates. This strategy is known to work in
compressed sensing
\cite{KrzakalaMezard12,KrzakalaPRX2012,DonohoJavanmard11} so that its
application in the present context should be straightforward. Another
interesting direction would be to replace the random matrices used in
this work by structured ones allowing a fast multiplication, as it has been
done, again, in compressed sensing
\cite{JavanmardMontanari12,barbier2013compressed}, in order to have a
decoder as fast as those used in sparse codes. We plan to study these
points in a future work.

\section*{Acknowledgment}
The research leading to these results has received funding from the
European Research Council under the European Union's $7^{th}$
Framework Programme (FP/2007-2013)/ERC Grant Agreement 307087-SPARCS
and from the French Ministry of defense/DGA.

\bibliographystyle{IEEEtran}
\bibliography{IEEEabrv,refs}

\begin{thebibliography}{10}
\providecommand{\url}[1]{#1}
\csname url@samestyle\endcsname
\providecommand{\newblock}{\relax}
\providecommand{\bibinfo}[2]{#2}
\providecommand{\BIBentrySTDinterwordspacing}{\spaceskip=0pt\relax}
\providecommand{\BIBentryALTinterwordstretchfactor}{4}
\providecommand{\BIBentryALTinterwordspacing}{\spaceskip=\fontdimen2\font plus
\BIBentryALTinterwordstretchfactor\fontdimen3\font minus
  \fontdimen4\font\relax}
\providecommand{\BIBforeignlanguage}[2]{{%
\expandafter\ifx\csname l@#1\endcsname\relax
\typeout{** WARNING: IEEEtran.bst: No hyphenation pattern has been}%
\typeout{** loaded for the language `#1'. Using the pattern for}%
\typeout{** the default language instead.}%
\else
\language=\csname l@#1\endcsname
\fi
#2}}
\providecommand{\BIBdecl}{\relax}
\BIBdecl

\bibitem{barron2010sparse}
A.~R. Barron and A.~Joseph, ``Sparse superposition codes: Fast and reliable at
  rates approaching capacity with gaussian noise,'' \emph{Manuscript. Available
  at “http://www. stat. yale. edu/~ arb4}, 2010.

\bibitem{joseph2012least}
A.~Joseph and A.~R. Barron, ``Least squares superposition codes of moderate
  dictionary size are reliable at rates up to capacity,'' \emph{Information
  Theory, IEEE Transactions on}, vol.~58, no.~5, pp. 2541--2557, 2012.

\bibitem{DonohoMaleki09}
D.~L. Donoho, A.~Maleki, and A.~Montanari, ``Message-passing algorithms for
  compressed sensing,'' \emph{Proc. Natl. Acad. Sci.}, vol. 106, no.~45, pp.
  18\,914--18\,919, 2009.

\bibitem{BayatiMontanari10}
M.~Bayati and A.~Montanari, ``The dynamics of message passing on dense graphs,
  with applications to compressed sensing,'' \emph{IEEE Transactions on
  Information Theory}, vol.~57, no.~2, pp. 764 --785, 2011.

\bibitem{Rangan10b}
S.~Rangan, ``Generalized approximate message passing for estimation with random
  linear mixing,'' in \emph{IEEE International Symposium on Information Theory
  Proceedings (ISIT)}, 2011, pp. 2168 --2172.

\bibitem{KrzakalaPRX2012}
F.~Krzakala, M.~M{\'e}zard, F.~Sausset, Y.~Sun, and L.~Zdeborov{\'a},
  ``Statistical physics-based reconstruction in compressed sensing,''
  \emph{Phys. Rev. X}, vol.~2, p. 021005, 2012.

\bibitem{RichardsonUrbanke08}
T.~Richardson and R.~Urbanke, \emph{{Modern Coding Theory}}.\hskip 1em plus
  0.5em minus 0.4em\relax Cambridge University Press, 2008.

\bibitem{MezardParisi87b}
M.~M{\'e}zard, G.~Parisi, and M.~A. Virasoro, \emph{Spin-Glass Theory and
  Beyond}.\hskip 1em plus 0.5em minus 0.4em\relax Singapore: World Scientific,
  1987, vol.~9.

\bibitem{MezardMontanari09}
M.~M\'ezard and A.~Montanari, \emph{Information, Physics, and
  Computation}.\hskip 1em plus 0.5em minus 0.4em\relax Oxford: Oxford Press,
  2009.

\bibitem{barron2011analysis}
A.~R. Barron and A.~Joseph, ``Analysis of fast sparse superposition codes,'' in
  \emph{Information Theory Proceedings (ISIT), 2011 IEEE International
  Symposium on}.\hskip 1em plus 0.5em minus 0.4em\relax IEEE, 2011, pp.
  1772--1776.

\bibitem{montanari2012graphical}
A.~Montanari, ``Graphical models concepts in compressed sensing,''
  \emph{Compressed Sensing: Theory and Applications}, pp. 394--438, 2012.

\bibitem{KrzakalaMezard12}
F.~Krzakala, M.~M\'ezard, F.~Sausset, Y.~Sun, and L.~Zdeborov\'a,
  ``Probabilistic reconstruction in compressed sensing: Algorithms, phase
  diagrams, and threshold achieving matrices,'' \emph{J. Stat. Mech.}, 2012.

\bibitem{Tanaka02}
T.~Tanaka, ``A statistical-mechanics approach to large-system analysis of cdma
  multiuser detectors,'' \emph{IEEE Trans. Infor. Theory}, vol.~48, pp.
  2888--2910, 2002.

\bibitem{KabashimaWadayama09}
Y.~Kabashima, T.~Wadayama, and T.~Tanaka, ``A typical reconstruction limit of
  compressed sensing based on lp-norm minimization,'' \emph{J. Stat. Mech.}, p.
  L09003, 2009.

\bibitem{NishimoriBook}
H.~Nishimori, \emph{Statistical Physics of Spin Glasses and Information
  Processing}.\hskip 1em plus 0.5em minus 0.4em\relax Oxford: Oxford University
  Press, 2001.

\bibitem{barron2012high}
A.~R. Barron and S.~Cho, ``High-rate sparse superposition codes with
  iteratively optimal estimates,'' in \emph{Information Theory Proceedings
  (ISIT), 2012 IEEE International Symposium on}.\hskip 1em plus 0.5em minus
  0.4em\relax IEEE, 2012, pp. 120--124.

\bibitem{KudekarRichardson12}
S.~Kudekar, T.~Richardson, and R.~Urbanke, ``Spatially coupled ensembles
  universally achieve capacity under belief propagation,'' 2012,
  arXiv:1201.2999v1 [cs.IT].

\bibitem{KudekarRichardson10}
------, ``Threshold saturation via spatial coupling: Why convolutional ldpc
  ensembles perform so well over the bec,'' in \emph{Information Theory
  Proceedings (ISIT),}, 2010, pp. 684--688.

\bibitem{YedlaJian12}
A.~Yedla, Y.-Y. Jian, P.~S. Nguyen, and H.~D. Pfister, ``A simple proof of
  threshold saturation for coupled scalar recursions,'' 2012, arXiv:1204.5703v1
  [cs.IT].

\bibitem{DonohoJavanmard11}
D.~L. Donoho, A.~Javanmard, and A.~Montanari, ``Information-theoretically
  optimal compressed sensing via spatial coupling and approximate message
  passing,'' in \emph{Proc. of the IEEE Int. Symposium on Information Theory
  (ISIT)}, 2012.

\bibitem{JavanmardMontanari12}
A.~Javanmard and A.~Montanari, ``Subsampling at information theoretically
  optimal rates,'' in \emph{Information Theory Proceedings (ISIT), 2012 IEEE
  International Symposium on}, July 2012, pp. 2431--2435.

\bibitem{barbier2013compressed}
J.~Barbier, F.~Krzakala, and C.~Sch{\"u}lke, ``Compressed sensing and
  approximate message passing with spatially-coupled fourier and hadamard
  matrices,'' \emph{arXiv preprint arXiv:1312.1740}, 2013.

\end{thebibliography}

\end{document}